\documentstyle[12pt]{article}
\newcommand{\be}{\begin{equation}}
\newcommand{\ee}{\end{equation}}

\newcommand{\bear}{\begin{eqnarray}}
\newcommand{\eear}{\end{eqnarray}}

\newcommand{\mod}{\left|\hspace{-0.4cm}\begxin{array}{c}\\ 
\end{array}\right.}

\renewcommand\({\left(}
\renewcommand\){\right)}
\renewcommand\[{\left[}

\newcommand\eq[1]{Eq.~(\ref{#1})}
\newcommand\eqs[2]{Eqs.~(\ref{#1}) and (\ref{#2})}

\newcommand\eqsss[4]{Eqs.~(\ref{#1}), (\ref{#2}), (\ref{#3})
and (\ref{#4})}

\newcommand\eea{\end{eqnarray}}
\newcommand\bea{\begin{eqnarray}}
        		
\newcommand\TeV{\,\mbox{TeV}}
\newcommand\GeV{\,\mbox{GeV}}
\newcommand\MeV{\,\mbox{MeV}}
\newcommand\keV{\,\mbox{keV}}
\newcommand\eV{\,\mbox{eV}}

\newcommand\mpl{M_{\rm Pl}}

\newcommand\luv{\Lambda_{\rm UV}}

\newcommand\lsim{\mathrel{\rlap{\lower4pt\hbox{\hskip1pt$\sim$}}
    \raise1pt\hbox{$<$}}}
\newcommand\gsim{\mathrel{\rlap{\lower4pt\hbox{\hskip1pt$\sim$}}
    \raise1pt\hbox{$>$}}}

\newcommand{\dlabel}{\label}

\def\dslash{\not{\hbox{\kern-2pt $\partial$}}}
\def\Dslash{\not{\hbox{\kern-4pt $D$}}}
\def\Oslash{\not{\hbox{\kern-4pt $O$}}}
\def\Qslash{\not{\hbox{\kern-4pt $Q$}}}
\def\pslash{\not{\hbox{\kern-2.3pt $p$}}}
\def\kslash{\not{\hbox{\kern-2.3pt $k$}}}
\def\qslash{\not{\hbox{\kern-2.3pt $q$}}}

 \newtoks\slashfraction
 \slashfraction={.13}
 \def\slash#1{\setbox0\hbox{$ #1 $}
 \setbox0\hbox to \the\slashfraction\wd0{\hss \box0}/\box0 }
 
\newcommand\sub[1]{_{\rm #1}}

\newcommand\vpq{F\sub{PQ}}
\newcommand\trh{T\sub{RH}}

\newcommand\oma{\Omega\sub a}

\newcommand\ax{\rm a}

\begin{document}

\begin{titlepage}

\title{\bf The abundance of relativistic axions in a flaton
model of Peccei-Quinn symmetry}

\author{ 
{\bf Eung Jin Chun$^{a,b}$, Denis Comelli$^b$ and David H. Lyth$^a$} \\[1ex]
$^a$Department of Physics, Lancaster University\\
Lancaster LA1 4YB. U.K. \\[1ex]
$^b$Korea Institute for Advanced Study\\ 
Seoul 130-012, Korea  \\[1ex]
$^c$INFN, sezione di Ferrara\\
Ferrara, via Paradiso 12 44100, Italy\\[1ex]
}


\maketitle
\def\baselinestretch{1.15}
\begin{abstract}
Flaton models of Peccei-Quinn symmetry have good 
particle physics motivation, and are likely to cause thermal inflation
leading to a well-defined cosmology.  They can solve the $\mu$ problem, 
and generate viable neutrino masses. 
Canonical flaton models
predict an axion decay constant $\vpq\sim 10^{10}\GeV$ and generic
flaton  models give $\vpq\gsim 10^9\GeV$ as required by observation.
The axion is a good candidate for cold dark matter in all cases,
because its density is diluted by flaton decay if $\vpq\gsim 10^{12}\GeV$.
In addition to the dark matter axions, a population of relativistic 
axions is produced by flaton decay, which at nucleosynthesis is equivalent
to some number  $\delta N_\nu$ of
extra neutrino species.
Focussing on the canonical model, containing three flaton particles
and two flatinos, we evaluate all of  the flaton-flatino-axion interactions
and the corresponding axionic decay rates. They are compared with the dominant
hadronic decay rates, for both DFSZ and KSVZ models. 
These formulas provide the basis for a precise  calculation of  the 
equivalent $\delta N_\nu$ in terms of the parameters (masses and couplings). 
The  KSVZ case is probably already  ruled out by the existing bound
$\delta N_\nu\lsim 1$. The 
 DFSZ case is allowed in a significant region of parameter space,
and will provide a possible explanation for any future detection of
nonzero $\delta N_\nu$.

\end{abstract}

\thispagestyle{empty}

\end{titlepage}

\def\baselinestretch{1.1}

\section{Introduction}

With the discovery of the instantons it was realized that the 
$\theta\sub{QCD}$ parameter of the Standard Model 
can have important physical consequences. In particular,
the induced CP 
violation  affects the electric dipole moment of the neutron, leading to 
the upper limit limit $\theta\sub{QCD}\leq 10^{-10}$. 
An attractive explanation for such a small parameter 
is provided by the Peccei-Quinn mechanism \cite{peccei,kim}.
There is supposed to be a spontaneously broken global $U(1)$ symmetry 
(PQ symmetry), which is also explicitly broken by the color anomaly.
Its pseudo-Goldstone boson is the axion.
The PQ symmetry is spontaneously broken by 
some set of scalar fields $\phi_i$ (elementary or composite)
with charges
$Q_i$, so that their PQ transformation is
\be
\phi_i \to e^{iQ_i \alpha} \phi_i \,.
\dlabel{pqsym}
\ee
Denoting the vacuum expectation value $\langle |\phi_i|\rangle$ by
$v_i/\sqrt 2$, we define the PQ symmetry breaking scale $\vpq$ by
\be
\vpq^2=\sum_i Q_i^2v_i^2
\,.
\ee
(In defining $\vpq$, we use the canonical normalization of the PQ charges,
that the smallest $Q_i^2$ is set equal to 1.)
Collider and astrophysics constraints require roughly \cite{raffelt}
\be 
\vpq/N \gsim  10^9 \GeV \dlabel{vpqastroph}
\,.
\ee
The bound is actually one on the axion mass, whose relation to $\vpq$ is 
given by
\be
m\sub a = 6 N\times 10^{-4}
\frac{10^{10}\GeV}{\vpq} \eV
\dlabel{amass}
\,.
\ee
Here,  $N$ is the number of distinct vacua, or equivalently the number of domain
walls meeting at each PQ string.

PQ charge will also be carried by 
fields which do not spontaneously
break PQ symmetry.
We shall consider KSVZ (hadronic)  models \cite{ksvz} in which
these are only some extra heavy quark superfields, and
DFSZ \cite{dfsz} models in which they are only
Standard Model (SM) superfields.

We are concerned\footnote
{An earlier version of this work appeared as hep-ph/9903286.}
 with models in which the
fields breaking PQ symmetry are flatons. Flatons are fields
whose tree-level potential
is flat in the limit of unbroken renormalizable supersymmetry,
which acquire nonzero vevs after soft supersymmetry breaking.
We make the usual assumption,  that the supersymmetry breaking is
  gravity-mediated.

In the rest of this section we recall the essential features of
flaton and non-flaton models of PQ symmetry.
In the next section we 
discuss in some detail the cosmology of flaton models.
In Section \ref{s2}, we give the  general
structure of the flaton  and flatino  masses.
In Section \ref{s3} we analyze 
the general self interactions between flatons and flatinos.
In Section \ref{s4} we 
see the effect of the interaction of the flatons with the matter fields.
In Section \ref{s5} we consider the implication of the nucleosynthesis bound 
on the energy density of relativistic axions, in particular, of a possible
future  bound $\delta N_\nu<0.1$.
We conclude in Section \ref{s6}.

\subsection{Models of PQ symmetry-breaking}

Let us consider the potential of the fields $\phi_i$
which break PQ symmetry. In a supersymmetric model there have to be at 
least two, but for first orientation we 
pretend that there is only 
one. In the limit of exact PQ  symmetry, its potential will be 
of the form
\be
V=V_0-m^2 |\phi|^2+\frac14 \lambda|\phi|^4 +
\sum_{n=1}^\infty
 \lambda_n \frac{|\phi|^{2 n +4}}{\mpl^{2n}}
\,.
\dlabel{phipot}
\ee
$\mpl=(8\pi G_N)^{-1/2}
=2.4\times 10^{18}\GeV$ is the reduced Planck mass.

\subsubsection{Non-flaton models}

If the renormalizable coupling $\lambda$ is of order
$1$, the non-renormalizable terms are negligible
and the mass of the radial oscillation is
$m\sim \vpq$. (Recall that $\vpq
=\sqrt2 Q \langle |\phi|\rangle$.)

Going on to the case where a number of fields $\phi_i$ ($i=1,\cdots,p$)
break PQ symmetry,
there will be $p$ particles corresponding to the radial
oscillations, and (in addition to the axion) $p-1$ 
particles corresponding to the angular oscillations.
There will also be $p$ superpartners with spin $1/2$.
In a non-supersymmetric theory all of these particles would
have mass of order $\vpq$, but supersymmetry protects the
mass of one scalar and one spin-$1/2$ particle.

Indeed, in the limit of unbroken supersymmetry, the holomorphy of the 
superpotential ensures that PQ symmetry 
is accompanied by a symmetry acting on the radial parts of the
PQ charged fields
\be
\phi_i\to e^{Q_i \alpha} \phi_i \,.
\dlabel{saxsym}
\ee
The corresponding pseudo-Goldstone boson is called the saxion
(or saxino), and the spin-$1/2$ partner of the axion/saxion 
is called the axino.
With gravity-mediated supersymmetry breaking, the saxion and
axino will typically both have soft masses of order $100\GeV$,
although specific models exist \cite{chunlukas}
with an axino mass of order $\keV$. In gauge-mediated models,
the saxion and axino will typically both have the sub-$\keV$ mass.

In non-flaton models one can hope to understand a
value $\vpq\sim 10^{10}\GeV$ since that is the supersymmetry 
breaking scale \cite{kim84}, but it may be hard to understand a bigger 
value.

\subsubsection{Flaton models}

We are concerned with models 
\cite{yamamoto,lps,cr,hitoshi1,hitoshi2}
in which the fields breaking PQ symmetry are flaton fields
\cite{flatons}.
This means that their tree-level potential is flat in  the limit of unbroken,
renormalizable supersymmetry. We assume that 
 supersymmetry breaking is gravity-mediated, which is usual in the context
of flaton fields.

Considering first the case of one flaton with potential
\eq{phipot}, the 
quartic term is (practically) zero, while 
the mass $m$ of the flaton 
is of order $100\GeV$ to $1\TeV$.
When making numerical estimates, we shall take for definiteness
\be
m=10^{2.5\pm 0.5}\GeV
\,,
\dlabel{mest}
\ee
The leading non-renormalizable term
generates a large vacuum expectation value (vev),
\be
\vpq=\sqrt 2
\langle|\phi|\rangle \sim
\( m\mpl^n/\sqrt{\lambda_n} \)^\frac1{n+1} \,.
\dlabel{vevexp}
\ee
Here, $\lambda_n$ is the coefficient of the leading term in
\eq{phipot}, which is expected to be roughly of order 1. In the 
case $n=1$ this gives a vev of order $10^{10}\GeV$, but it can
be bigger if $n$ is bigger. 
Imposing the condition that $V$ (practically) vanishes in
the vacuum gives the
height of the potential,
\be
\(\frac{V_0^{1/4}}{10^6\GeV}\) \sim 
\(\frac{\langle|\phi|\rangle}{10^{10}\GeV} \)^{1/2} \,.
\ee

We are concerned with the case that
the $\phi_i$ are flaton fields.
The scalar particles are now the axion, plus $2p-1$ flatons 
with mass of order $100\GeV$, and $p$ flatinos with masses
of the same order. 
The saxion, defined through
\eq{saxsym}, is a linear combination of flatons,
while the axino
(defined as the partner of the
axion-plus-saxion) is a linear combination of flatino 
mass eigenstates. Neither of them has any
special significance. In particular, 
the possibility of a
$\keV$-mass axino does not exist in flaton models.
In the models that we shall consider, $p=2$ so that there are three 
flatons and two flatinos.

\subsection{Estimates of $\vpq$}

\dlabel{svpq}

To estimate the magnitude of $\vpq$ in flaton models,
we begin with schematic case of one flaton field, with potential
\eq{phipot}.

Consider first the expected magnitude of the coefficients
$\lambda_n$ in \eq{phipot}.
In the case that $\mpl$ represents the ultra-violet cutoff for the
effective
field theory, one usually assumes
$\lambda_n\sim 1$, but
$\lambda_n\sim 1/(2n+4)!$ may be more realistic
\cite{km}. On the other hand, one might
quite reasonably suppose that the cutoff, call it $\luv$,
is around the gauge coupling
unification scale $10^{-2}\mpl$
(either because fields of a Grand
Unified Theory have been integrated out, or because this is the
true quantum gravity scale). In that case, the estimate
$\lambda_n\sim 1/(2n+4)!$ to 1 would be reasonable if
$\mpl$ were changed   
to $\luv$ in \eq{phipot}. Retaining $\mpl$, one should
multiply the estimate of
$\lambda_n$ by a factor $(\mpl/\luv)^{2n}\sim 10^{4n}$.  
In view of these considerations, we adopt as a reference
the estimate $1/(2n+4)!\lsim \lambda_n\lsim
10^{4n}$, corresponding to
$\lambda_1^\frac 14=10^{0.2\pm 0.8}$ and
$\lambda_2^\frac 16 =10^{0.3\pm1.0}$.

Using these estimates of $\lambda_n$, we can make
estimates of $\vpq$ bearing in mind the uncertainty
\eq{mest} in $m$. In these and other estimates, we  
add in quadrature different uncertainties
in the exponents. This procedure has no particular basis,
but at least it is better than ignoring the uncertainties completely,
or adding different estimates linearly. In the case at hand, the
uncertainty is dominated by the large uncertainty that we assigned to
the $\lambda_n$. We take the PQ charge of $\phi$ to be 1,
so that $\vpq=\sqrt2\langle|\phi|\rangle$.

Unless it is forbidden by a symmetry, the leading term $n=1$
will be the one appearing in \eq{vevexp}, leading to
\be
\vpq=10^{10.4\pm 0.9}\GeV \hspace{4em} (n=1)
\,.
\dlabel{vpqest1}
\ee
If the leading term is $n=2$ this becomes
$\vpq= 10^{12.9\pm 1.1}\GeV$. At higher $n$, $\vpq$ slowly increases,
so for $n>1$ we have
\be
\vpq>10^{11.8}\GeV \hspace{4em} (n>1) \,.
\dlabel{vpqest2}
\ee
   
In the class of supersymmetric models that we shall consider,
PQ symmetry is broken by two flaton fields $\phi\sub P$ and
$\phi\sub Q$, with charges respectively $1$ and $2n+1$.
The fields interact, giving the rather complicated potential
\eq{spot} for $n=1$ and one of similar form \cite{cck} for
bigger $n$. In all cases, the vevs of both flaton fields
are given roughly by \eq{vevexp}. There are additional
uncertainties because there are more soft parameters and factors
involving $n$, but in view of the large uncertainty we already assigned
to the coupling $\lambda_n$ \eqs{vpqest1}{vpqest2} should still provide
reasonable estimates.

\section{Cosmology and dark matter}

\subsection{Cosmology of the PQ fields}

Generic models of
PQ symmetry have many
possible cosmological consequences, which 
have been discovered gradually over the years.

In all cases the
axion lifetime is longer than the age of the Universe
\cite{kim},
so that its present density
must be $\oma\leq 0.3$ or so, with the equality prevailing if
axions are the dark matter. 
The
density depends on 
whether the axions come from
strings 
or from the vacuum fluctuation of the axion field
during inflation. (In the latter case the axion density \cite{myaxion2}
depends on our  location in the universe.)
Strings may be produced by a variety of mechanisms during \cite{lyst}
and after inflation.
The axion density depends also
on the amount of any entropy production after the axion
mass switches on at a temperature around $1\GeV$.
As a result there is no model-independent prediction for 
$\oma$ as a function of $\vpq$.

Within the usual framework of non-flaton models, the 
superpartners of the axion can also have a range of cosmological
consequences. 
An axino with a $\keV$ mass is  a dark matter candidate,
which may be produced by a variety of mechanisms
and give rise to a variety of cosmological consequences
\cite{kevaxino,chunlukas}.
Alternatively, an axino with a $10\GeV$ mass may 
be the cold dark matter \cite{ckr}, as it can be the 
lightest supersymmetric particle (LSP).
Finally, the saxion is a late-decaying particle
which may be produced by thermal or other mechanisms. If it is 
sufficiently abundant 
to dominate the density of the Universe, it
must decay well before nucleosynthesis, and before it does so it 
will dilute the abundance pre-existing relics, including baryons and
dark matter candidates \cite{kimsax,mysax}. 
If it is less abundant it may decay much later, and affect the
formation of large-scale structure \cite{hangbae}.

In contrast with this generic situation, the cosmology of
flaton models is rather well-defined, on the reasonable assumption
\cite{thermal3,cck} that
the PQ flaton fields generate an era of thermal inflation
\cite{thermal1,thermal2,thermal3,thermal4,thermal5,thermal6},
which is not followed by any other such era.
Thermal inflation occurs long after the ordinary inflation which is 
supposed to be origin of structure, and 
may wipe out all previously existing
relics. 
When it ends, PQ strings are produced, and 
flaton decay (the analogue of saxion decay) produces a
calculable amount of entropy, with the reheating at a calculable 
temperature in the range $\MeV$ to $\TeV$.
As a result, 
the axion density is in principle calculable, and appears to be
compatible with the observed dark matter density
for any $\vpq$
allowed by other considerations (Section \ref{sdm}.)
In other words, the axion is a good dark matter candidate
in flaton models.
Finally, the flatinos
 (the generalizations 
of the axino) cannot have the $\keV$ mass, and for simplicity
 we assume
that none of them is the LSP. (The opposite case will be explored in a future
paper \cite{ckl}.)

A unique feature of flaton models is that flaton decay creates
a highly relativistic population of axions
\cite{thermal3,cck}.
This population has nothing to do with the dark matter.
Its density at nucleosynthesis is 
equivalent to roughly $\delta N_\nu\sim 1$ 
extra neutrino species \cite{cck}. 
Flaton models of PQ symmetry will therefore be a candidate for 
explaining a nonzero $\delta N_\nu$ that may be established 
in the future, and the models will be strongly constrained
if the present bound \cite{delnu}
$\delta N_\nu\lsim 1$ is significantly tightened.

\subsection{Thermal inflation and reheating}

In the early 
Universe, with Hubble parameter $H\gsim 100\GeV$, fields
with the true soft mass $|m|\sim 100\GeV$ are expected 
\cite{msquared}
to have an 
effective mass-squared $m^2(t)\sim
\pm H^2$. 
(During inflation this result
might be avoided \cite{treview}, 
but it should still hold 
afterwards \cite{drt,gtr,p99nont}.) This applies in particular to the
flaton fields.
Pretending for the moment that there is only one flaton field,
we  focus on the case that  $m^2(t)$ is positive in the early Universe,
because thermal inflation 
\cite{thermal1,thermal2,thermal3,thermal4,thermal5,thermal6}
then occurs, leading to rather definite predictions for the cosmology.

Let us summarize the history. After
$H$ falls below
$|m|$, there will be enough thermalization
to hold $\phi$ at the origin until
thermal inflation begins. (See Section IIIB of \cite{thermal3}.)
Thermal inflation begins when 
the potential $V_0$ dominates the energy density.
This is at the epoch $T\sim V_0^{1/4}\sim 10^6\GeV$
(assuming for simplicity that full reheating has occurred by that time).
Thermal inflation
ends after $\sim \ln\(V_0^{1/4}/|m|\)\lsim$ 10 $e$-folds, when the
temperature is of order $|m|\sim 100\GeV$.

When thermal inflation ends, the flaton field
$\phi$ moves away from the origin.
Cosmic strings form, and between them the roughly homogeneous
flaton field starts to oscillate around its vev.
Corresponding to the oscillation is a population of flatons.
We discuss in Section \ref{sparres} the possibility that
parametric resonance rapidly drains away the energy of this
oscillation, finding that this phenomenon will probably not occur
and will in any case have little effect on the following considerations.
Discounting parametric resonance, 
energy loss comes at first only
from the Hubble drag, which is negligible during one oscillation.
The oscillation  corresponds to flatons, 
with conserved number and non-relativistic random motion (matter as opposed
to radiation). When the flatons decay,
the Universe thermalizes, at the
reheat temperature \cite{thermal3}
\bea
T\sub{RH} &\simeq&  10^{0.3}g\sub{RH}^{-\frac14}
\sqrt{\mpl \Gamma} 
\simeq 10^{-0.2} \sqrt{10^{-2} \mpl c N\sub{chan} m^3 \vpq^{-2} }
      \dlabel{treh1} \\
&=& 10^{2.2\pm0.5} \(\frac{10^{10}\GeV}{\vpq} \) \( 
\frac{m}{10^{2.5}\GeV} \)^{3/2}
\GeV \,.
\dlabel{best}
\eea
In the first expression, $\Gamma$ is a typical flaton decay rate,
while $g\sub{RH}\sim 100$ is the effective number of particle species
at $T\sub{RH}$.
In the second expression $N\sub{chan}$ is the number of decay channels,
$c$ is a factor of order 1 and $m$ is a typical soft parameter.
The factor $10^{-2}$ in the estimate of $\Gamma$ is what one expects
in the case of unsupressed couplings for a single decay
channel \cite{thermal3}, and it is confirmed by, for instance,
the estimates in \eq{f2aa}, {\it etc}. There are in reality several channels,
and for definiteness,
we take $cN\sub{chan}=10^{1.0\pm 1.0}$, leading to the estimate in the 
third line. 

Adding in quadrature the uncertainty of \eq{mest}, we find
\be
\trh= 10^{2.2\pm 0.9} \(\frac{10^{10}\GeV}{\vpq} \) \GeV \,.
\ee
In order not to upset nucleosynthesis, it is  required to have
$\trh\gsim 10\MeV$, which corresponds to
\be
\vpq\lsim 10^{15}\GeV \,.
\dlabel{vpqbound}
\ee
Using \eqsss{vpqest1}{vpqest2}{mest}{best}, we estimate
\bea
T_{RH}&=& 10^{1.8\pm1.3}\GeV \hspace{4em}(n=1) 
\dlabel{test1}\\
T_{RH}&\lsim & 10^{0.4}\GeV \hspace{4em}(n>1)  \dlabel{test2}
\,.
\eea

In this discussion of thermal inflation and its aftermath, we have
retained the pretense that there is just one flaton field.
There are in the models we shall consider
two flaton fields $\phi_P$ and $\phi_Q$, with the potential \eq{spot}
or its $n>1$ analogue. We assume that $m\sub P^2(t)$
is positive in the early Universe, so that thermal inflation occurs.
When thermal inflation ends, $\phi\sub P$ moves away from the origin,
and as a result $\phi\sub Q$ also moves away from the origin.
At first the orbit
in field space will be far from the vev, but after a few Hubble times
the Hubble drag will allow the vev to attract the orbit towards it,
so that there are almost sinusoidal oscillations of the eigenmodes
around their vacuum values. These oscillations are equivalent to the
presence of the three species of flaton, and each of
them decays at the epoch specified by \eq{treh1}, with $m$
the appropriate mass. The reheating process is complete after
the last decay has taken place.

\subsection{Dark Matter and baryons}

\dlabel{sdm}

\subsubsection*{Axionic dark matter}

The axion number density is conserved after some epoch
$T\sub{cons}\sim 1\GeV$. In the case $n=1$, 
$\trh$ is bigger than $T\sub{cons}$ and entropy is conserved too.
The axion density is then expected to be of the form
\be
\oma = C\(\frac{\vpq}{10^{12}\GeV}\)^{1.2} \,.
\ee
The constant $C$ is in principle calculable from the dynamics of the 
strings, walls and axions, derived ultimately from the field equation
of the flaton fields breaking PQ symmetry.
According to one group \cite{ds,bs1,bs2}, 
$C\sim 1$ to $10$, while according to another
\cite{hs,hcs1,hcs2,hcs3}, $C\sim 0.2$. 
In view of this uncertainty, which we emphasize
is one of computation rather than principle, we conclude that
the axion is a good  dark
matter candidate in the case $n=1$ which corresponds to \eq{vpqest1}.
(By a `good' candidate, we mean one whose density is predicted to 
be within at least a few orders of magnitude of the observed dark
matter density.)

In the case $n>1$, $\trh$ is smaller than $T\sub{cons}$.
Entropy is produced until the epoch $\trh$, giving \cite{cck,mynew}
\be
\oma = \tilde C \(\frac{10^{12}\GeV}{\vpq} \)^{0.44} \,,
\ee
with roughly  $\tilde C\sim 10C$. At least if $C$ is not
too big, the axion is a good dark matter candidate in these
models too \cite{cck}.

\subsubsection*{Baryogenesis and the LSP}

If thermal inflation wipes out pre-existing relics, baryogenesis
has to occur after thermal inflation.
The crucial factor here is the final reheat temperature
$T\sub{RH}$. 
Baryogenesis mechanisms occurring at the electroweak phase transition
can operate if 
$T\sub{RH}\gsim 100\GeV$. 
This is possible in
the canonical case $n=1$, but not in the case $n>1$.
If $\trh$ is smaller one must turn to other mechanisms,
which are quite speculative. Proposals include
a complicated Affleck-Dine mechanism along the lines of
\cite{thermal5}, QCD baryogenesis \cite{robert} or parametric resonance
baryogenesis \cite{parambar}. 

If  the lightest supersymmetric particle (LSP)
is stable,  it has to thermalize in order to
avoid overproduction from flaton decay. This requires
$T\sub{RH}\gsim m\sub{LSP}/20$ \cite{cck}, say $T\sub{RH}$
more than a few GeV. This is exactly what one expects in 
the case $n=1$. As is well known, a stable LSP is a good dark matter
candidate.

If the LSP is unstable (due to $R$-parity violation), 
baryogenesis can occur simply by allowing a
baryon-number violating flaton decay
channel (the Dimopoulos-Hall mechanism \cite{dh}).
This mechanism requires DSFZ as opposed to KSVZ coupling to matter,
and as we  shall see the former case is favored in flaton models.
For the mechanism to work, final 
reheat must occur at a temperature less than a few GeV 
\cite{dh}. This is likely for $n>1$, but looks rather unlikely
in the canonical case $n=1$.

Let us summarize. In the canonical case $n=1$, 
the LSP can thermalize, and therefore can be stable
so that it is a good dark matter candidate just like the axion. In this case, 
baryogenesis  mechanisms involving 
the electroweak phase transition can operate.
In the case $n>1$, the LSP cannot thermalize and 
therefore cannot be stable. 
Baryogenesis from flaton decay (the Dimopoulos-Hall mechanism)
is a natural possibility in this case.

\subsubsection*{Supermassive dark matter}

We have seen that the axion is a good dark matter candidate,
and that in the canonical model the LSP is also a good candidate.
These conclusions hold both in the DFSZ and KSVZ cases.
In the  KSVZ case, there is a third  good dark matter, namely
the heavy quarks $E, E^c$, which are  strongly interacting massive
particles (SIMPs).
Until thermal inflation ends, the  SIMPs are light and will be in thermal
equilibrium. After thermal inflation ends,
SIMPs acquire mass
through the coupling 
 $\phi E E^c$ to the flaton. The density of such particles  has been shown 
\cite{thermal6} to be naturally in the right ballpark. 

The  scenario of \cite{thermal6}
should be contrasted with an earlier proposal \cite{tonietal},
that the supermassive particle is very heavy
also in the early Universe, and never in thermal 
equilibrium. Such a particle may be produced by the vacuum
fluctuation during ordinary
inflation \cite{tonietal}, or by other mechanisms. 
In the former case,
each comoving wavenumber leaving the horizon during inflation will
acquire very roughly one particle per quantum state
\cite{withdave,lrs,grt}, leading
very roughly to \cite{lrs,tonietal,grt}
\be
\Omega_0 = \(\frac{m}{10^{14}\GeV} \) \(\frac{H_*}{10^{14}\GeV}\)
\(\frac{T\sub{INFRH}}{10^9\GeV} \) \gamma
\,.
\ee
In this expression, $m$ is the mass of the superheavy particle,
$H_*$ is the Hubble parameter during slow-roll inflation, and $T\sub{INFRH}$
is the temperature at reheat after slow-roll inflation, and $\gamma$ 
is the dilution caused by entropy production after that epoch.
In our case, thermal
inflation will give roughly $\gamma\sim e^{-10}$.
One can adjust the other parameters to make $\Omega_0\sim 1$,
but in contrast
with the LSP and the axion the required value $\Omega_0\sim 1$
is not favored over any other. In other words,
this kind of supermassive dark candidate is not (in our present state
of knowledge) a {\em good} dark matter candidate, merely
a possible one.\footnote
{The same is true in the case $\gamma=1$. In particular,
the choice $m\sim H_*$ advocated in \cite{tonietal} is not
in fact particular favored, bearing in mind that $m$ is the true mass
as opposed to the effective mass during inflation. 
Inflation with a potential $V=\frac12 m\sub{INF}^2\phi\sub{INF}^2$
indeed ends when $H\sim m\sub{INF}$, but the inflaton field
$\phi\sub{INF}$ is not supposed to be stable and hence is not
a dark matter candidate.}

\subsection{Parametric resonance?}

\dlabel{sparres}

To check whether parametric resonance \cite{param1} occurs,
we make the very crude approximation that the 
sinusoidal oscillation corresponding to the three flatons 
is present from the very beginning. We also assume that the 
masses of the three flatons are roughly the same, or else
that one of the amplitudes is much bigger than the others. 
Then the
field equation of the Fourier component of each produced field
$\phi_n$ is a Mathieu equation, leading to a situation that has been 
analyzed in the literature \cite{param1}.
(More realistic cases,
including the one where the oscillation starts at a maximum of the
potential \cite{green}, seem to give similar results.)
The oscillation of the real field
$\phi_I$ corresponding to the $I$th flaton 
leads to an oscillating mass-squared $m^2_n(\phi_I)$
for each of the produced scalar fields.
Parametric resonance occurs, leading to
significant production of $\phi_n$, if
\cite{tkachev}
\be
q \equiv m_{0n}^2(\phi_I)/m^2 \gsim 10^3 \,.
\ee
In this formula, $m\sim 100\GeV$ is a typical oscillation frequency,
and $m^2_{0n}(\phi_I)$ is the amplitude of $m^2_n(\phi_I)$.
The initial amplitude of oscillation
is $\phi_{I0}\sim \vpq$.
The scalar particles that can be produced include 
the flatons and the axion,
and in  the DFSZ case also
the
Standard Model Higgs and sfermions. From Sections \ref{s3} and \ref{s4},
each of these has $m_{0n}\sim 100\GeV$, making $q\sim 1$.
The conclusion is that parametric 
resonance probably does not occur, but a more detailed 
calculation is needed to say anything definite especially in view
of the extremely anharmonic nature of the initial oscillation.

Even if it occurs, the overall effect of
parametric resonance will not be dramatic, unless it
leads to baryogenesis \cite{parambar}.
Its initial effect is to 
quickly damp the flaton field oscillation, producing flatons,
axions and in the DFSZ case Higgs and sfermions.
The produced particles have very roughly the same energy density,
and are  marginally relativistic except for the axions 
which are highly relativistic. The flatons are stable on the Hubble
timescale while the Higgs and sfermions decay into highly relativistic
ordinary matter (plus the marginally relativistic LSP if it is stable). 
After a small number of Hubble times the energy in the
highly relativistic particles becomes negligible. The
dominant energy is in non-relativistic flatons,
coming partly from the parametric resonance and partly from the 
residual homogeneous 
oscillation of the flaton
fields. 
Except for the baryogenesis possibility, there is no change.

\subsection{Relativistic axions}

\dlabel{srelax}

Now we come to the relativistic axions, which will be our 
concern for the rest of the paper.
This axion population comes from 
the decay of the flatons \cite{thermal3,chunlukas}
when they finally reheat the Universe.
Its density during nucleosynthesis is conveniently specified
by the equivalent number of extra neutrino species
\be
 \delta N_{\nu} \equiv \(\frac{\rho\sub a}{\rho_{\nu}}\)_{NS}
\,,\dlabel{neutrino}
\ee
where  $\rho\sub a$ is the energy density of relativistic axions, and 
  $\rho_{\nu}$ is 
 the energy density of a single species of relativistic neutrino.

At present, constraints coming from nucleosynthesis are bedeviled
by the fact that there are two separate allowed regions of parameter
space, corresponding to  `low' and `high' deuterium densities.
In the `high' region, the  bound \cite{delnu}
at something like
2-$\sigma$ level is  $\Delta N_\nu< 1.8$, while  in the perhaps
favored  `low' region,  the bound at a similar level is
$\Delta N_\nu < 0.3$. As we shall see, the flaton models
predict $\delta N_\nu$ roughly of order 1, so that at least the
second bound is quite constraining.

In the canonical model that we shall discuss, there are three flaton
species, and in more general models there are more flatons. In general,
each  flaton species can decay into relativistic  hadronic matter
${\rm X}$, into ${\rm X}\ax$ where $\ax$ denotes a relativistic axion,
or into $\ax\ax$. (We neglect for simplicity the
 small branching ratio into  channels containing more axions.)
Let us pretend  first that there is one flaton $\phi$.
The hadrons ${\rm X}$  thermalize
immediately, but the axions do not thermalize
\cite{cck}, so their
density at reheating is 
\be
\rho\sub a =
B\sub a \rho\sub r
\,,
\ee
where
\be
B\sub a\equiv \frac{\Gamma(\phi\to \ax\ax)+\frac12\Gamma(I\to {\rm X}\ax)}
{\Gamma(\phi\to {\rm X})}
\,,
\nonumber
\ee
and the density of thermalized radiation is
$$
\rho\sub r =\frac{\pi^2}{30} g\sub{RH} T\sub{RH}^4
\,.
$$
After reheating, $\rho\sub a\propto (a\sub{RH}/a)^4
=(s\sub{NS}/s\sub{RH})^\frac43$,
where $a$ is the scale factor, and
$s$ is the entropy density of the particles in thermal
equilibrium. The latter is given by
$s=(2\pi^2/45)gT^3$, where $g$ is the effective number of relativistic
species
in thermal equilibrium and $T$ is their temperature. At
the beginning of the 
nucleosynthesis era, $g=10.75$ and
\be
\rho\sub a= B\sub a \frac{\pi^2}{30} g\sub{RH}
\(\frac{10.75}{g\sub{RH}}\)^\frac43 T\sub{NS}^4 
\,.
\ee
The density of one neutrino species at that epoch is
$$
\rho_\nu = \frac{\pi^2}{30} \frac{8}{7} T\sub{NS}^4
\,,
$$
so that 
\be
\delta N_\nu \equiv
13.6 g\sub{RH}^{-\frac13} B\sub a  
= 2.9  B\sub a  
\dlabel{delax}
\,.
\ee
For the last equality  we used 
$g\sub{RH} =100$, appropriate if  $T\sub{RH}\sim1\TeV$.
The  alternative choice $g\sub{RH}\simeq 10$ would reduce
$\delta N_\nu$ by a factor 2 or so.

In models with more flatons, denoted by a label $I$, the
quantity $B\sub a$ to be used in \eq{delax} is
given by
\be
B\sub a\equiv
\frac{\sum_I n_I \( \Gamma(I\to \ax\ax)+ \frac12 \Gamma(I\to {\rm X}\ax)\)}
{\sum_I n_I\Gamma(I\to {\rm X})}
\,,\dlabel{badef}
\ee
where   $n_I$ is
 the number density of the flaton $I$ just before reheating.

If the flaton fields suffered negligible energy loss
until they start their sinusoidal oscillation about
their vev, one could
in principle calculate the $n_I$ by solving the field equation of motion
under the potential \eq{spot}. The same thing is possible if the energy
loss can be calculated, the only known paradigm 
for that purpose being the parametric resonance approximation.
Such a calculation would be difficult, and its  uncertainty
 impossible to quantify at present. 

One can however say something useful without knowing the $n_I$,
by considering the quantities
\be
B_I\equiv  \frac{\Gamma(I\to \ax\ax)+
\frac12(I\to {\rm X}\ax)}{\Gamma(I\to {\rm X})}
\,.
\ee
If all of these quantities were equal, they would be equal to $B\sub a$.
 More usefully,
if they are all known to have an upper or a lower bound $r$,
in some regime of parameter space, then
$B\sub a$ has the same bound.
In the case of an upper bound, one concludes from \eq{delax}
that $\delta N_\nu < 4.4r$, making the model in this regime
{\em compatible} with 
a given bound on $\delta N_\nu$ if $r$ is small enough. In the case of an upper
bound, one concludes that $\delta N_\nu >4.4r$, making the model in this 
regime {\em incompatible} with a given bound if $r$ is big enough.

\section{Flaton and flatino spectrum}

\dlabel{s2}

\subsection{The superpotential}

The model we consider contains
two flaton superfields $\hat P$ and $\hat Q$,
interacting with the superpotential \cite{hitoshi1,hitoshi2} 
\be
W\sub{flaton}=\frac{f}{\mpl^n}\hat{ P}^{n+2}\hat{ Q} 
\dlabel{Wflaton}
\,.
\ee
We deal with the simplest case $n=1$ and assign the PQ charges $-1$ and $3$
to $\hat P$ and $\hat Q$, respectively.
Here we note that  quantum gravity  may break  PQ symmetry,
giving  nonrenormalizable terms which invalidate the PQ solution to
the strong CP problem. 
  A way to avoid this is to impose a certain discrete gauge 
symmetry forbidding sufficiently higher dimensional operators \cite{disgau}.
However, to have such a discrete symmetry, one has to extend the model 
beyond the
simple superpotential (\ref{Wflaton}) under consideration.
  Our analysis can be
applied to such extended cases with a straightforward generalization.

With the inclusion of the  soft
susy breaking terms  and the cosmological constant, the potential is
\bear
V&=& V_0 + m_P^2 |\phi\sub P|^2+
m_Q^2|\phi\sub Q|^2+
\frac{ f^2}{\mpl^2}  \(9  |\phi\sub P|^4 |\phi\sub Q|^2 +
|\phi\sub P|^6\)+
\(\frac{A_f}{\mpl} f \phi\sub P^3\phi\sub Q+h.c.\) \,.
\dlabel{spot}
\eear
The soft parameters $m_P$, $m_Q$ and $A_f$ are all of order
$100\GeV$ in magnitude.
It is assumed that $m_P^2$ and $m_Q^2$ are both positive at the Planck 
scale. 
The interactions of $\phi_P$ with the right handed neutrino
superfields  give radiative corrections
which drive  $m_P^2$ to a negative value  at the PQ scale, 
generating vevs $v_P$ and $v_Q$ for respectively $|\phi_P|$
and $|\phi_Q|$. According to \eqs{vpqest1}{vpqest2},
both vevs are roughly
of order $10^{10}\GeV$.
As we shall discuss in Section \ref{s2}, the radial oscillations of
the flaton fields $\phi_P$ and $\phi_Q$ correspond to
two flatons, while the angular oscillations correspond
to a third flaton and the axion.

\subsection{Flaton spectrum}

We write the flaton fields as
\bea
\phi\sub P &=& \frac{v_P+P}{\sqrt{2}}  e^{i \frac{A_P}{v_P}} \nonumber\\
\phi\sub Q &=& \frac{v_Q+Q}{\sqrt{2}}  e^{i \frac{A_Q}{v_Q}} \,.
\eea
{}From now on, we shall take $v_P$ and $v_Q$ as the independent 
parameters trading with $m^2_P$ and $m^2_Q$ in the potential \eq{spot}.
The vevs are taken to real and positive, and we shall
take the other independent paramerters to be $A_f$ and $f$ 
to be real with opposite sign.

The axion field is 
\be
a= -\frac{v_P}{F_{PQ}} A_P +3 \frac{v_Q}{F_{PQ}} A_Q 
\ee
where $F_{\rm PQ}^2=v_P^2+9v_Q^2$.
The orthogonal field to the axion (both are CP Odd)  corresponds to a 
flaton. It is
\be
\psi'=
-\frac{v_P}{F_{\rm PQ}} A_Q -
3 \frac{v_Q}{F_{\rm PQ}} A_P \,.
\ee
With our choice
$A_f f<0$, the vev is at $\psi'=0$, and the mass-squared is
\be
M_{\psi'}^2=-\frac{f A_f v_P F_{\rm PQ}^2}{2  \mpl v_Q}
=-\frac{f}{g} \mu A_f \( x^2+9\) 
\ee
where 
\bea
\frac{\mu}{g} &\equiv& \frac{ v_P v_Q}{2 \mpl} 
\dlabel{gmu} \\
x &\equiv & \frac{v_P}{v_Q} \,.
\eea
For future convenience we have introduced a
quantity $\mu$, related to the $g$ appearing 
only in the DFSZ model. At this stage results
depend only on the ratio $\mu/g$ defined by (\ref{gmu})
and they apply to both models.

The other two flatons correspond to the
CP even fields $P$ and $Q$. They have a 2$\otimes$2 mass matrix whose
components are 
\bear
M^2_{QQ}&=& M_{\psi'}^2 \frac{x^2}{9+x^2} \\\nonumber
M^2_{PQ}&=& 9  f^2 \frac{ v_P^4}{ \mpl^2 x}- 
3\frac{ M_{\psi'}^2 x}{ 9+x^2}=
3 x\(12  \frac{f^2}{g^2} \mu^2  - 
 \frac{ M_{\psi'}^2 }{ 9+x^2}\)
 \\\nonumber
M^2_{PP}&=&3  f^2 \frac{v_P^4\(x^2+3 \)}{ \mpl^2 x^2 }-
3 \frac{M_{\psi'}^2 }{9+x^2}=
12  \frac{f^2}{g^2} \(x^2+3 \) \mu^2 -
3 \frac{M_{\psi'}^2 }{9+x^2} \,.
\\\nonumber
\eear
Here two mass parameters $m_P^2, m_Q^2$ in Eq.~(\ref{spot}) are replaced 
in favor of $v_P, v_Q$.
Performing the rotation from the flavor basis $||P\; Q||$ to the mass basis
$||F_1\; F_2||$
\bea
 P &=& \cos\alpha F_2 -\sin\alpha F_1 \nonumber \\
 Q &=& \sin\alpha F_2 + \cos\alpha F_1  \,,
\eea
we find the mixing angle $\alpha$ determined by
%
\bea
 \cos2\alpha &=& \frac{M^2_{PP}-M^2_{QQ}}{M^2_{F_2}-M^2_{F_1}}
     = \epsilon \frac{x^2+3}{\sqrt{9+42x^2+x^4}}  \nonumber\\
 \sin2\alpha &=& \frac{2M^2_{PQ}}{M^2_{F_2}-M^2_{F_1}}
     = \epsilon \frac{6x}{\sqrt{9+42x^2+x^4}} 
\dlabel{cs2a}
\eea
where $\epsilon\equiv{\rm sign}M^2_{PQ}$ as 
we have the relation, $M^2_{F_2}-M^2_{F_1}=
|M^2_{PQ}/3x|\sqrt{9+42x^2+x^4}$.
Later, the decay rates can be expressed in terms of 
$\cos2\alpha$ and $\sin2\alpha$ without ambiguity in fixing the
angle $\alpha$ itself.
The two eigenstates $F_1, F_2$ have masses,
\be
M^2_{F_{2,1}}=\frac{\mu^2}{2}\left(
\frac{f}{g}(12(x^2+3) \frac{f}{g}+
(3-x^2)\frac{A_f}{\mu})\pm
|\frac{f}{g} (12 \frac{f}{g}+
\frac{A_f}{\mu}) |\sqrt{
9+42 x^2 +x^4}\right) \,
\dlabel{flagen}
\ee
with $M_{F_2} > M_{F_1}$.  

The requirement $M^2_{F_1}>0$ gives the constraint  
\be
y_1<y=-\frac{g \;A_f}{f\; \mu}
\frac{9+ x^2}{ 4 x^2}<y_2 \,,
\ee
where
$$
y_{1,2}\equiv \frac{9+ x^2}{ 8  x^2}\(21+ x^2 \pm 
\sqrt{9+42  x^2+ x^4}\)
$$
or
\be \dlabel{f1bound}
 \frac{1}{2}\(21+ x^2 -
\sqrt{9+42 x^2+ x^4}\)<-\frac{g \;A_f}{f\; \mu}<\frac{1}{2}\(21+ x^2 +
\sqrt{9+42  x^2+ x^4}\) \,.
\ee
One can find $M_{F_1}<M_{\psi'}$ for all the parameter space, and thus
$F_1$ is the lightest flaton.


\subsection{Flatino spectrum  }

{}From the   superpotential $W\sub{flaton}$
we can directly extract   also the flatino's mass matrix
whose eigenvalues are
\be
M^2_{\tilde{F}_{2,1}}=\frac{9}{4} \frac{M^2_{\psi'}}{y\;  x^2}[ x^2 +2 \pm 
 2 \sqrt{x^2+1}]=9\frac{f^2}{g^2}\mu^2[x^2 +2 \pm 
 2 \sqrt{x^2+1}] 
\dlabel{flatinomass}
\ee
The eigenstates $\tilde{F}_1, \tilde{F}_2$ are related to the 
flavor states $\tilde{P}, \tilde{Q}$ by 
\bear
\tilde{F}_1 &=& \cos \tilde{\alpha} \tilde{P}+ \sin  \tilde{\alpha}
\tilde{Q}  \nonumber\\
\tilde{F}_2 &=& -\sin  \tilde{\alpha}
 \tilde{P} +  \cos \tilde{\alpha} \tilde{Q} 
\dlabel{angle}
\eear
where 
the angle $ \tilde{\alpha}$ is determined by 
\bea
\cos2\tilde{\alpha}&=&-\frac{1}{\sqrt{1+x^2}} \nonumber\\
\sin2\tilde{\alpha}&=&-\frac{x}{\sqrt{1+x^2}}  \,.
\eea
A parameter space analysis indicates that  we have always 
$M_{F_1} \leq  2 M_{\tilde{F}_{1}}$.
 This automatically  forbids  the decay of $F_1$ to 
flatinos
leaving  open only the decay into flatinos of the heavier 
 $F_2$ and  $\psi'$  flatons.

\section{Decays involving only flatons, flatinos and axions}
\dlabel{s3}


In this section, we analyze the various decay rates between flatonic
fields. We begin with the decay channels induced by the kinetic
term and the superpotential $W_{\rm flaton}$, which are common to
the KSVZ and DFSZ models.

\subsection{ Derivative  and cubic interaction terms between flatons}

The flaton interaction terms with at least one derivative are given by the 
Lagrangian 
\bear 
L_{\partial}&=&
\frac{2\, v_P \, P+P^2}{2 v_P^2} [
\frac{v_P^2}{F_{\rm PQ}^2} \(\partial a\)^2
+9\frac{v_Q^2}{F_{\rm PQ}^2} \(\partial \psi'\)^2 +
6 \frac{v_Q v_P}{F_{\rm PQ}^2} \partial \psi' \partial a]+ \nonumber\\
&&\frac{2\, v_Q \, Q+Q^2}{2 v_Q^2} [ 
9 \frac{v_Q^2}{F_{\rm PQ}^2} \(\partial a\)^2
+\frac{v_P^2}{F_{\rm PQ}^2} \(\partial \psi'\)^2 -
6 \frac{v_Q v_P}{F_{\rm PQ}^2} \partial \psi' \partial a]
\dlabel{derivas}\eear
{}From this expression we can extract the following terms expressed 
in mass eigenstates:

(i) the trilinear derivative interactions with no axions,
\be
\(\partial \psi'\)^2\frac{1}{F_{\rm PQ}  x \sqrt{x^2+9}}
[\(-9\sin \alpha + x^3 \cos \alpha\) F_1 +
\(9 \cos \alpha + x^3 \sin \alpha\)F_2] \,;
\dlabel{psis}
\ee

(ii) the  trilinear derivative interactions with only one axion,
\be
L_{ F_i \psi'a}=  \partial \psi' \partial a  \frac{6} 
{F_{\rm PQ}\, \sqrt{ x^2+9}}[
 \(\cos \alpha - x \sin \alpha\)F_2-\(\sin \alpha +  x \cos \alpha\)F_1] \,;
\ee

(iii) the  trilinear derivative interactions with  two axions,
\be
L_{F_i a a}  =
|\partial_{\mu}a|^2   \frac{1}{F_{\rm PQ} \sqrt{ x^2+9}}\( \(9 \cos \alpha-x\,
 \sin \alpha\)F_1+ \(9\, \sin \alpha + x\, \cos \alpha\) F_2\) \,.
\ee
All the above derivative interactions can be 
transformed in scalar interactions if  we are working at tree level and with
on-shell  external particles
\be
\phi\sub 1 \(\partial_{\mu} \phi\sub 2\) \(\partial^{\mu} \phi\sub 3\)=
\frac{1}{2} \(M_{\phi\sub 1 
}^2-M_{\phi\sub 2}^2-M_{\phi\sub 3}^2\)\phi\sub 1\phi\sub 2\phi\sub 3
\ee
The cubic interactions come also from the 
superpotential and the soft terms
\begin{eqnarray} 
L_{\phi^3}&=&\({9\over2}f^2
 {v_P v_Q^2 \over \mpl^2} +{5\over2}f^2{v_P^3 \over
 \mpl^2} -
    \frac{ M_{\psi'}^2 v_Q x}{v_P^2 \(x^2+9\)} \) P^3   
+ \({27\over2}f^2{v_P^2 v_Q\over \mpl^2}
-3\frac{ M_{\psi'}^2  x}{v_P \(x^2+9\)} \)P^2Q
\nonumber \\
&+& \({9\over2}f^2{v_P^3\over \mpl^2}\) PQ^2
+ \({3f\over4}{A_fF_{\rm PQ}^2 \over \mpl v_Q}\)P\psi'\psi'
+ \({f\over4}{A_fF_{\rm PQ}^2v_P \over \mpl v_Q^2}\)Q\psi'\psi' \,.
\dlabel{cubis}\end{eqnarray}
In the mass basis, we get
\bear L_{F_2F_1F_1}&=&\nonumber
\left\{ -3\frac{M_{\psi'}^2\,\sin \alpha}
{F_{\rm PQ}x \sqrt{9 + x^2}} 
\,\( -2 x\cos^2 \alpha + 
        x\,\sin^2\alpha +
        \,\cos\alpha\,\sin\alpha 
 \) + \right.
 \\\nonumber 
 &&
 \frac{6\;f^2\,\mu^2 \sqrt{x^2+9}}
{g^2 x F_{\rm PQ}} \( -18\,\cos^2\alpha\,\sin\alpha \,x +
     9\,\sin^3\alpha\,x + 3 \,\cos^3\alpha \,x^2 \right. \\
&& \left.\left. -\cos\alpha\, \sin^2\alpha (-9 + x^2) \)    \right\}
F_1^2 F_2 \equiv \frac{A_{F_2F_1F_1}}{2}F_1^2 F_2
\eear
For the  full trilinear $F_2 \psi'\psi'$ interaction, we have to add up
the terms in Eqs.~(\ref{psis}) and (\ref{cubis}) to obtain
%
%
\bear
L_{F_2 \psi'\psi'}=
\frac{ M_{F_2}^2 }{2 x \sqrt{9+x^2}F_{\rm PQ}}
\(-\(3 \sin \alpha +x \cos \alpha\)\(x^2+9\)
\frac{M_{\psi'}^2}{M_{F_2}^2}+\right.
\\\nonumber
\left.
\(9 \cos \alpha +\sin \alpha x^3\)\(1-2\frac{M_{\psi'}^2}{M_{F_2}^2}\)\)
F_2 \psi'\psi'\equiv \frac{A_{F_2 \psi'\psi' }}{2}\psi'^2 F_2
\eear

Collecting the above formulae one finds the  decay rates 
among flatons and axions  

\begin{eqnarray}
&&\Gamma\(F_2\to aa\) = {1\over 32\pi} {M^3_{F_2}\over F^2_{\rm PQ} \(x^2+9\)
 }  \(x\cos\alpha + 9 \sin\alpha\)^2 \label{f2aa}\\
&&\Gamma\(F_1\to aa\) = {1\over 32\pi} {M^3_{F_1}\over F^2_{\rm PQ} \(x^2+9\)
 }       \(-x\sin\alpha + 9 \cos\alpha\)^2 \dlabel{f1aa}\\
&&\Gamma\(F_2\to F_1F_1\) =
\frac{1}{32 \pi M_{F_2} }\sqrt{1-4 \frac{M^2_{F_1}}{M^2_{F_2}}}
|A_{F_2F_1F_1}|^2 \\
&&\Gamma\(F_2\to \psi'\psi'\) = 
\frac{1}{32 \pi M_{F_2} }\sqrt{1-4 \frac{M^2_{\psi'}}{M^2_{F_2}}}
|A_{F_2\psi'\psi'}|^2 \\
&&\Gamma\(F_2\to a\psi'\) = {1\over 16\pi} {M^3_{F_2}\over F^2_{\rm PQ} 
   \(x^2+9\) } \(1-{M^2_{\psi'} \over M^2_{F_2}}\)^{3}
              \(3\cos\alpha-3x\sin\alpha\)^2 \\ 
&&\Gamma\(\psi'\to aF_2\) = 
  {1\over 16\pi} {M^3_{\psi'}\over F^2_{\rm PQ}
   \(x^2+9\) } \(1-{M^2_{F_2} \over M^2_{\psi'}}\)^{3}
              \(3\cos\alpha-3x\sin\alpha\)^2 \\
&&\Gamma\(\psi'\to aF_1\) = 
    {1\over 16\pi} {M^3_{\psi'}\over F^2_{\rm PQ}
   \(x^2+9\) } \(1-{M^2_{F_1} \over M^2_{\psi'}}\)^{3}
              \(3\sin\alpha+3x\cos\alpha\)^2 
\end{eqnarray}
Energy conservation will of course forbid some of these reactions,
depending on the flaton masses. 
As  $M_{F_1}<M_{\psi'}$ 
the channels $F_1\rightarrow \psi'\psi'$ and
$F_1\rightarrow \psi'a$ are always forbidden.

\subsection{  Interaction terms between flatons  and flatinos }

The trilinear Lagrangian terms responsible for the decay of flatons or 
flatinos are   
\bear 
L_{\phi  \bar{\tilde{\phi}} \tilde{\phi}}&=&
\frac{3 f}{2\mpl}\(\(v_PQ+v_QP\)\bar{ \tilde{P}} \tilde{P}+
i \frac{v_P^2+3 v_Q^2}{F_{\rm PQ}} \psi' \bar{
\tilde{P}}\gamma_5 \tilde{P}
-i2 \frac{v_P v_Q}{F_{\rm PQ}} a \bar{\tilde{P}}\gamma_5 \tilde{P}\)
+ \nonumber\\
&& \frac{3 f}{2 \mpl}\(2 P v_P\bar{ \tilde{P}} \tilde{Q} +
2i \frac{v_P}{F_{\rm PQ}} \(3 v_Q \psi'+v_P a\)\bar{\tilde{P}}\gamma_5\tilde{Q}
\)
\dlabel{yukas} \eear
(the tilded fields are the fermionic
 superpartner of the respective $P$ and $Q$ scalars).
Let us denote the Yukawa couplings between  the flaton (or the axion) and 
the flatinos in mass basis 
by $-L_{Yuk}=Y_{ijk} \phi_i \bar{\tilde{F}}_j\(1,\gamma_5\)
\tilde{F}_k/2$ where $\gamma_5$ is taken for $\phi_i=a, \psi'$.
We find from Eq.~(\ref{yukas}) the following expressions for the
Yukawa couplings 
\bear
&& Y_{F_1\tilde{F}_1\tilde{F}_1}= {6f\mu\sqrt{x^2+9} \over gxF_{\rm PQ} }
[\(x\cos\alpha-\sin\alpha\)\cos^2\tilde{\alpha}-x\sin\alpha\sin2\tilde{\alpha}]
  \nonumber \\
&& Y_{F_1\tilde{F}_2\tilde{F}_2}= {6f\mu\sqrt{x^2+9} \over gxF_{\rm PQ} }
[\(x\cos\alpha-\sin\alpha\)\sin^2\tilde{\alpha}+x\sin\alpha\sin2\tilde{\alpha}]
  \nonumber \\
&& Y_{F_1\tilde{F}_1\tilde{F}_2}= {6f\mu\sqrt{x^2+9} \over gxF_{\rm PQ} }
[-x\sin\alpha\cos2\tilde{\alpha}+{1\over2}\(\sin\alpha-x\cos\alpha\)
  \sin2\tilde{\alpha}]      \nonumber \\
&& Y_{F_2\tilde{F}_1\tilde{F}_1}= {6f\mu\sqrt{x^2+9} \over gxF_{\rm PQ} }
[\(x\sin\alpha+\cos\alpha\)\cos^2\tilde{\alpha}+x\cos\alpha\sin2\tilde{\alpha}]
  \nonumber \\
&& Y_{F_2\tilde{F}_2\tilde{F}_2}= {6f\mu\sqrt{x^2+9} \over gxF_{\rm PQ} }
[\(x\sin\alpha+\cos\alpha\)\sin^2\tilde{\alpha}-x\cos\alpha\sin2\tilde{\alpha}]
  \nonumber \\
&& Y_{F_2\tilde{F}_1\tilde{F}_2}= {6f\mu\sqrt{x^2+9} \over gxF_{\rm PQ} }
[+x\cos\alpha\cos2\tilde{\alpha}-{1\over2}\(\cos\alpha+x\sin\alpha\)
  \sin2\tilde{\alpha}]      \nonumber \\
&& Y_{\psi'\tilde{F}_1\tilde{F}_1}= {6f\mu \over gxF_{\rm PQ} }
[-\(3+x^2\)\cos^2\tilde{\alpha}-3x\sin2\tilde{\alpha}]    \nonumber       \\
&& Y_{\psi'\tilde{F}_2\tilde{F}_2}= {6f\mu \over gxF_{\rm PQ} }
[-\(3+x^2\)\sin^2\tilde{\alpha}+3x\sin2\tilde{\alpha}] \nonumber \\
&& Y_{\psi'\tilde{F}_1\tilde{F}_2}= {6f\mu \over gxF_{\rm PQ} }
[-3x\cos2\tilde{\alpha}+{1\over2}\(3+x^2\)\sin2\tilde{\alpha}] \nonumber \\
&& Y_{a\tilde{F}_1\tilde{F}_1}= {6f\mu \over gxF_{\rm PQ} }
[2x\cos^2\tilde{\alpha}-x^2\sin2\tilde{\alpha}] \nonumber \\
&& Y_{a\tilde{F}_2\tilde{F}_2}= {6f\mu \over gxF_{\rm PQ} }
[2x\sin^2\tilde{\alpha}+x^2\sin2\tilde{\alpha}] \nonumber \\
&& Y_{a\tilde{F}_1\tilde{F}_2}= {6f\mu \over gxF_{\rm PQ} }
[-x^2\cos2\tilde{\alpha}-x\sin2\tilde{\alpha}] 
\eear
{}From this we can extract the decay rates for
$F_i\to \tilde{F_j}\tilde{F_k}$,
$\psi'\to \tilde{F_j}\tilde{F_k}$, or
$\tilde{F_2} \to \tilde{F_1} F_i \(\psi',a\)$ 
\begin{eqnarray}
\Gamma\(F_i \to \tilde{F_j}\tilde{F_k}\) &=&
   {M_{F_i} \over 8\pi } S
   \( 1-{(M_{\tilde{F_j}}+M_{\tilde{F_k}})^2\over M^2_{F_i}} \)^{3\over2}
\( 1-{(M_{\tilde{F_j}}-M_{\tilde{F_k}})^2\over M^2_{F_i}} \)^{1\over2}
   Y^2_{F_i\tilde{F_j}\tilde{F_k}}  \\
\Gamma\(\psi'\to \tilde{F_j}\tilde{F_k}\) &=&
   {M_{\psi'} \over 8\pi } S 
   \(1-{(M_{\tilde{F_j}}+M_{\tilde{F_k}})^2\over M^2_{\psi'}} \)^{1\over2}
   \(1-{(M_{\tilde{F_j}}-M_{\tilde{F_k}})^2\over M^2_{\psi'}} \)^{3\over2}
   Y^2_{\psi'\tilde{F_j}\tilde{F_k}}  \\
\Gamma\(\tilde{F_2}\to \tilde{F_1}F_i\) &=&
   {M_{\tilde{F_2}} \over 16\pi } 
   \(1-{(M_{\tilde{F_1}}+M_{{F_i}})^2\over M^2_{\tilde{F_2}}}\)^{1\over2}
   \(1-{(M_{\tilde{F_1}}-M_{{F_i}})^2\over M^2_{\tilde{F_2}}}\)^{1\over2} 
\nonumber\\
&\times&
   \( \(1+{M_{\tilde{F_1}} \over M_{\tilde{F_2}}}\)^2 -
          {M^2_{F_i} \over M^2_{\tilde{F_2}}} \)
   Y^2_{F_i\tilde{F_1}\tilde{F_2}}  \\
\Gamma\(\tilde{F_2}\to \tilde{F_1}\psi'\) &=&
   {M_{\tilde{F_2}} \over 16\pi } 
   \(1-{(M_{\tilde{F_1}}+M_{\psi'})^2\over M^2_{\tilde{F_2}}}\)^{1\over2}
   \(1-{(M_{\tilde{F_1}}-M_{\psi'})^2\over M^2_{\tilde{F_2}}}\)^{1\over2} 
\nonumber\\
&\times&
   \( \(1-{M_{\tilde{F_1}} \over M_{\tilde{F_2}}}\)^2 -
          {M^2_{\psi'} \over M^2_{\tilde{F_2}}} \)
   Y^2_{\psi'\tilde{F_1}\tilde{F_2}}  \\
\Gamma\(\tilde{F_2}\to \tilde{F_1}a\) &= &
   {M_{\tilde{F_2}} \over 16\pi } 
   \(1-{M_{\tilde{F_1}}^2\over M^2_{\tilde{F_2}}}\)
   \( 1-{M_{\tilde{F_1}} \over M_{\tilde{F_2}}}\)^2 
   Y^2_{a\tilde{F_1}\tilde{F_2}}  
\end{eqnarray}
where $S$ is a symmetric factor ($1/2$ for identical final states or otherwise
 $1$).

\section{Interaction of flatons and flatinos with matter fields}
\dlabel{s4}

Now we study the interactions of the flatons with 
 matter  and supermatter.
Through these interactions the 
flatons and flatinos decay into ordinary matter and axions,
and the production of the latter must be
sufficiently suppressed to satisfy the
nucleosynthesis limit on the effective  number $\delta N_\nu$
 of extra neutrino species.
(In this paper, we are for simplicity assuming that no
 flatino is stable.)

\subsection{KSVZ model: Interactions between flatons  and gluons }

In the KSVZ (hadronic) model, the interaction with matter
is 
\be \dlabel{hadron}
W\sub{flaton-matter}=h_{E_i} \hat{ E}_i\hat{ E}^c_i\hat{ P}
\ee
where $\hat E_i$ and $\hat E_i^c$ are 
additional heavy quark and antiquark superfields.

The only decay mode available
for the flatons is into two gluons coming from the anomaly 
(when the space phase will be available, we have to take into account  also
the decay into  massive gluinos, 
in this discussion we neglect such a possibility).
The respective one loop corrected  decay rates are
\bea
\Gamma \(F_1 \rightarrow g+g\)&=&  \frac{\alpha_S^2\(M_{F_1}\)}{ 72 \pi^3}\,
N^2_E \, \frac{M_{F_1}^3}{x^2\;F_{\rm PQ}^2}\(x^2+9\) \sin^2\alpha
\(1+\frac{95}{4}\frac{\alpha_S\(M_{F_1}\)}{ \pi}\) \nonumber\\
\Gamma \(F_2 \rightarrow g+g\)&=& 
\frac{\alpha_S^2\(M_{F_2}\)}{ 72 \pi^3}\,N^2_E\,
\frac{M_{F_2}^3}     {x^2\;F_{\rm PQ}^2}\(x^2+9\) \cos^2\alpha
\(1+\frac{95}{4}\frac{\alpha_S\(M_{F_2}\)}{ \pi}\)
\label{glgl}
\eea
where $N_E$ is the total number of the
superheavy exotic quark fields ($M_E=h_E v_P \gg M_{F_i}$).
We do not consider the flatino decay into a gluon and a gluino which
will be irrelevant for our discussion.

\subsection{DFSZ model:
 Interactions between flatons/flatinos and ordinary matter}

In the DFSZ model, the interaction is
\be\dlabel{peccei}
W\sub{flaton-matter}
=\frac{1}{2} \lambda \hat{ N}\hat{ N}\hat{ P}+\frac{g}{\mpl}  \hat{ H}_1 
\hat{ H}_2 
\hat{ P} \hat{ Q}
\ee
where $ \hat{N}$ are the right handed neutrino superfields and
$\hat{H}_{1,2}$ the two Higgs doublets.  Due to the second term we can
provide a solution to the $\mu$ problem \cite{muprob}.
In such  case we can add to the superpotential of 
the minimal supersymmetric standard model also the terms
$h_{\nu}\hat{l}\hat{H_2}\hat{N}$
that generate the necessary mixing between left and right neutrinos to
implement a see--saw mechanism which can explain the solar 
and atmospheric neutrino deficits.

The decay properties of the flatons now involve the 
direct interactions between flatons and ordinary matter 
and supermatter 
In general the interaction between flatons and Higgs fields
are quite interesting  due to the fact that these two sectors,
after the spontaneous breaking of the PQ and the EW symmetry, mix together.
We notice that the Peccei-Quinn symmetry prevents the introduction of a  
SUSY invariant mass term $\mu H_1H_2$,  solving automatically   
the so--called $\mu$ mass problem as mentioned before.

Let us start by writing the Higgs-flaton potential
\begin{eqnarray} 
V\(H,\phi\) &=&
 \left| H_1 \right|^2
\( m_{H_1}^2 + \left| g{\phi\sub P \phi\sub Q \over \mpl} \right|^2 \)
+ \left| H_2 \right|^2
\( m_{H_2}^2 + \left| g{\phi\sub P \phi\sub Q \over \mpl} \right|^2 \)
 \nonumber  \\
 &&
+ \left\{ g H_1 H_2 \( A_g{\phi\sub P\phi\sub Q \over \mpl} +
   3f^*{\phi\sub P^{*2}|\phi\sub Q|^2 \over \mpl^2} +
   f^*{\phi\sub P^{*2}|\phi\sub P|^2 \over \mpl^2} \) +
 \mbox{c.c.} \right\} \\
&&
 + {1\over8} \(g^2+ {g'}^2\) \( |H_1|^2 - |H_2|^2 \)^2  \nonumber \,.
\dlabel{VHf}
\end{eqnarray}
When the fields $\phi\sub {P,Q}$ get  vevs, the  
$m_3^2H_1H_2$ mass term is generated dynamically.
The size of such a term is fixed  by
\be
m_3^2=
\mu\(A_g  +\frac{f}{g} \mu \(x^2+3\)\)
\ee
In the limit  $|m_3^2| \gg M_W^2$ the masses of the pseudoscalar $A^0$, of
the CP even scalar Higgs field $H^0$
and of the charged Higgs fields $H^{\pm}$
are almost degenerate
\be
M_{A^0,H^0,H^{\pm}}^2 \simeq -\frac{ m_3^2}{\sin \beta \cos \beta}
\ee
so from  the constraint of positivity of such a masses we get
\be \dlabel{pseudo}
\frac{A_g}{\mu}+\frac{f}{g}(x^2+3) \leq 0
\ee
In such a limit we also know that the mass eigenstate of the CP even 
electroweak sector $H^0,h^0$ and of the CP odd one $A^0,G^0$  are
\bear
H^0&=& -\sin \beta \;  h_1^0 +\cos \beta\; h_2^0 \nonumber \\
h^0&=& \cos \beta \;h_1^0 +\sin \beta \;h_2^0 \\
A^0&=& \sin \beta \;A_1^0 +\cos \beta \;A_2^0  \nonumber\\
G^0&=& \cos \beta \;A_1^0 -\sin \beta \;A_2^0  \nonumber 
\eear
where $H_1=\frac{1}{\sqrt{2}}\( v_1+h_1^0+iA_1^0\)$ and 
$ H_2= \frac{1}{\sqrt{2}}\( v_2+h_2^0+iA_2^0\)$  are the gauge eigenstates
and $\tan\beta=v_2/v_1$.
To allow the flaton decay into $A^0$,
we want it to be light so that small $\tan\beta$ is preferred in our discussion.
Hereafter we will take $\tan\beta=1$.

{}From Eq.~(\ref{VHf}),  we find 
\begin{eqnarray}
&&V_{Fhh}={1\over2} \mu^2 \(h_1^{02}+h_2^{02}+ A_1^{02}+A_2^{02}\)
                \({P \over v_P}+ {Q\over v_Q}\)
+  {1\over2} \left[\(h_1^0h_2^0-A_1^0A_2^0\) +
                    i \(h_1^0A_2^0+ h_2^0A_1^0\)\right] \nonumber\\
&&\left[ A_g\mu\({P\over v_P} + {Q \over v_Q} - i {x^2+3 \over x F_{\rm PQ}} 
   \psi'\) + 6 {f\over g} \mu^2 \({P \over v_P} + {Q \over v_Q} + 
  i{3\over xF_{\rm PQ}}\psi'\)
  + x^2{f\over g} \mu^2 \( 4{P \over v_P} + i{6\over xF_{\rm PQ}}\psi'\) \right]
  \nonumber\\
  && + c.c. 
\end{eqnarray}
It is then simple manner to get  the decay rates for 
the kinematically  more favorable decay channels
 $F_{1,2} \rightarrow h^0h^0$ and $\psi \to h^0 A^0$
\begin{eqnarray} 
&&\Gamma\(F_1\to h^0h^0\)=  
   {M^3_{F_1} \over 32\pi F_{\rm PQ}^2} \frac{\(x^2+9\)}{16\, x^2}
 {\mu^4\over M^4_{F_1}}
   \(1-{4M_{h^0}^2\over M_{F_1}^2}\)^{1/2} |A_{F_1 h h}|^2  \nonumber \\
&&\Gamma\(F_2\to h^0h^0\)=
   {M^3_{F_2} \over 32\pi F_{\rm PQ}^2}  \frac{\(x^2+9\)}{16\, x^2} 
{\mu^4\over M^4_{F_2}}
   \(1-{4M_{h^0}^2\over M_{F_2}^2}\)^{1/2}  |A_{F_2 h h}|^2 \\
&&\Gamma\(\psi'\to h^0A^0\)=  
  {M^3_{\psi'} \over 16\pi F_{\rm PQ}^2} {\mu^4\over M^4_{\psi'} }
\(1-{\(M_{h^0}-M_{A^0}\)^2\over M_{\psi'}^2}\)^{1/2}
  \(1-{\(M_{h^0}+M_{A^0}\)^2\over M_{\psi'}^2}\)^{1/2}
|A_{\psi h A}|^2 \nonumber
\dlabel{Fhh}
\end{eqnarray}
where 
\begin{eqnarray}
&& A_{F_1 hh} =\sin2\beta [ \({A_g\over \mu}
        + 6{f\over g}\)
      \(x\cos\alpha -\sin\alpha\)-4x^2{f\over g}\sin\alpha]
       + 2\(x\cos\alpha-\sin\alpha\)   \nonumber \\
&& A_{F_2 hh} = \sin2\beta [ \({A_g\over \mu} + 6{f\over g}\)
             \(x\sin\alpha +\cos\alpha\)+4x^2{f\over g}\cos\alpha]
            + 2\(x\sin\alpha+\cos\alpha\)  \nonumber\\
&& A_{\psi' hA} = \({A_g\over \mu} - 6{f\over g}\) {\(x^2+3\) \over x} 
\end{eqnarray}

The flatino decay into ordinary particles comes from the superpotential
$W = {g\over \mpl} \hat{H_1}\hat{H_2}\hat{P}\hat{Q}$.
We find that the flatino decay into a Higgs and a Higgsino 
(more precisely, the lightest neutralino $\chi_1$)
has the rate;
\begin{equation}
 \Gamma\(\tilde{F_i} \to \chi_1 h^0\) = 
 {M^3_{\tilde{F_i}} \over 8\pi F_{\rm PQ}^2} {\mu^2\over M^2_{\tilde{F_i}} }
 \(x^2+9\)^2 C_{\tilde{F}_i}^2 
 \( 1- {\(M_{\chi_1}+M_{h^0}\)^2 \over M^2_{\tilde{F_i}} }\)^{1/2}
 \( \(1+{M_{\chi_1} \over M_{\tilde{F_i}}}\)^2-
           {M^2_{h^0} \over M^2_{\tilde{f_I}}} \)
\end{equation}
where 
$C_{\tilde{F}_1}= (\sin\tilde{\alpha}+x^{-1}\cos\tilde{\alpha}) N_{\chi_1}$ 
and 
$C_{\tilde{F}_2}= (\cos\tilde{\alpha}-x^{-1}\sin\tilde{\alpha}) N_{\chi_1}$.
Here $N_{\chi_1}$ denotes the fraction of lightest neutralino in Higgsinos.

Let us now consider the flaton decay into ordinary fermions or sfermions.
The mixing terms between flaton and Higgs fields
allow a direct tree level coupling (after full mass matrix diagonalization)
between the usual fermions and flatons.
Parameterizing such a mixing with the parameter $\theta_{FH}$  the effective 
flaton-fermion interaction is $ h_f \theta_{FH}$
so that the rate of decay is
\be
\Gamma \(F_i \rightarrow f+\bar{f}\)= 
 N_c \frac{h_f^2  \theta_{FH}^2 } {16 \pi }M_{F_i} \(1-4\frac{m_f^2}
{M_{F_i}^2}\)^{\frac32}
\ee
where $N_c$ is a color factor for the fermion $f$.
Since  $\theta_{FH}\simeq \(\frac{v_{EW}}{F_{PQ}}\)$
\be
\Gamma \(F_i \rightarrow f+\bar{f}\)/\Gamma \(F_i \rightarrow a+a\)\sim
h_f^2 v_{EW}^2/M_{F_i}^2 \sim m_f^2/M_{F_i}^2\lsim 1 \,.
\ee
 Therefore, the rate of the flaton decay into ordinary fermions
cannot be made sufficiently larger than  that into axions.

For the coupling between sfermions and flatons, we have two contributions.
One is a direct coupling coming from the scalar potential
\bear
&& V_{F\tilde{f}\tilde{f}} = {\mu \over \vpq} 
{\sqrt{x^2+9}\over x}{ v_1 \over \sqrt{2}}
 \( h_d  \tan \beta \tilde{D_L}^* \tilde{D_R}^*
+  h_e  \tan \beta \tilde{E_L}^* \tilde{E_R}^*
+  h_u   \tilde{U_L}^* \tilde{U_R}^*\)\\
&& \nonumber
\(F_1\(x \cos \alpha-\sin \alpha\)+ 
F_2\(\cos \alpha+x \sin \alpha\)\)+h.c.
\eear
where $\tilde{D}^*$ denote down-type squark, {\it etc}.

The other  arises  from  an indirect 
coupling induced by  the mixing between Higgs and flaton fields as
for the fermion case.
Taking in consideration the cubic soft A-terms we find
\be
V_{eff}=h_d A_d\, \theta_{F_i\, H_1}\, F_i \tilde{D_L} \tilde{D_R}
+  h_e A_e \, \theta_{F_i\, H_1}\, F_i  \tilde{E_L} \tilde{E_R}
+  h_u A_u \,\theta_{F_i\, H_2}\, F_i \tilde{U_L} \tilde{U_R}
\ee
so that effectively we have couplings of the size 
\be
G_{F\, sfermion}\sim  h_f (\mu+A_f){v\sub{EW} \over \vpq} 
\ee

%
Diagonalizing the sfermion mass matrix
we can write $\tilde{f}_R \tilde{f}_L=a_{11}\tilde{f}_1\tilde{f}_1 +
a_{22}\tilde{f}_2\tilde{f}_2 +a_{12}\tilde{f}_1\tilde{f}_2$
(where $a_{ii}\propto h_f$  so for $h_f \rightarrow 0$ we have 
$a_{12}  \rightarrow 1$).
Considering the decay of the light flaton we get
\be
\Gamma \(F_1 \rightarrow \tilde{f}_{i}+\tilde{f}_{j}\) \simeq 
N_c 
\frac{G_{F\, \tilde{f}}^2 }
{64 \pi\; M_{F_1}}
a_{ij}^2
\sqrt{ 1-4 \frac{m_{\tilde{f}}^2 }{M_{F_1}^2}} 
\ee
As observed in Ref.~\cite{cck}, the flaton may decay efficiently to
two light stops as $h_t \sim 1$ and $a_{ij}\sim 1$ and thus
a large splitting between light and heavy stops helps increasing the
flaton decay rate to light stops. 
This kind of mass splitting occurs also in the Higgs sector and furthermore
the light Higgs ($h^o$)
 is usually substantially lighter than the heavy Higgs ($H^o$)
in  the minimal supersymmetric standard model.  This should be contrasted to
the case with the mass splitting for stops  which 
requires some adjustment in soft parameters.
In this  paper  we  concentrate on the flaton decay into 
Higgses, as the probably dominant mode, which in any case provides
a lower bound on the decay rates to ordinary matter and therefore an
upper bound on $\delta N_\nu$.

\section{ Parameter space  analysis}

\dlabel{s5}

\begin{table}
\caption{Direct decay channels involving only flatons
($I=F_1$, $F_2$ and $\psi'$), flatinos
($\tilde F_1$ and $\tilde F_2$), and the axion (a).
The only decays which must occur are
$F_i\to \ax\ax$, $\psi'\to \ax F_1$, and {\em either}
$\psi'\to \ax F_2$ {\em or} $F_2\to \psi'\ax$.
Any of the others may
be forbidden by energy conservation, if the 
decaying particle is too light.}
\begin{center}
\begin{tabular}{|lllll|}
\hline
$F_1$ & $F_2$ & $\psi'$ & $\tilde F_1$ & $\tilde F_2$ 
\\[4pt] \hline 
aa  & aa & $\ax F_1$ & {\rm none} & $\tilde F_1 I$ \\[4pt]
& $\ax\psi'$ & $\ax F_2$ && \\ [4pt]
& $\tilde F_i\tilde F_j$ & $\tilde F_i \tilde F_j$ & & \\[4pt]
& $F_1F_1$ &&& \\[4pt]
& $\psi'\psi'$ & & & \\[4pt]
\hline
\end{tabular}
\end{center}
\end{table}

We have now evaluated all of the 
direct decay rates for flatons and flatinos
into channels  involving axions, as summarized in Table 1.
We have also evaluated some of the contributions to the direct decay 
rates of flatons and flatinos into hadronic matter ${\rm X}$.
The ultimate objective is to evaluate the decay rates $\Gamma(I\to
\rm X)$, $\Gamma(I\to {\rm X}\ax)$, and $\Gamma(I\to \ax\ax)$ for each 
of the three flaton species, so as to evaluate $\delta N_\nu$
 through \eqs{delax}{badef}.
For the reactions 
  $F_i\to\ax\ax$ and $F_1\to \rm {\rm X}$, the direct rates
are the same as the total rates, but in the other cases one
has to consider also chain reactions.
 The reactions $F_2\to \rm {\rm X}$
and $\psi'\to \rm {\rm X}$ can go either directly or through
chains, and their rates are
\bea
\Gamma(\psi'\to {\rm {\rm X}}) &=& \Gamma\sub{dir}(\psi'\to {\rm {\rm X}})
+ \Gamma(\psi'\to \tilde F_1\tilde F_1) \nonumber\\
&&+ \Gamma(\psi'\to \tilde F_1\tilde F_2) B(\tilde F_2 \to {\rm {\rm X}})
+ \Gamma(\psi'\to \tilde F_2\tilde F_2) B^2(\tilde F_2\to {\rm {\rm X}}) \nonumber\\
\Gamma(F_2\to {\rm {\rm X}}) &=& \Gamma\sub{dir}(F_2\to {\rm {\rm X}})
+ \Gamma(F_2\to \tilde F_1\tilde F_1) \nonumber\\
&&+ \Gamma(F_2\to \tilde F_1\tilde F_2) B(\tilde F_2 \to {\rm {\rm X}})
+ \Gamma(F_2\to \tilde F_2\tilde F_2) B^2(\tilde F_2\to {\rm {\rm X}}) 
\,,
\eea
where $B$ denotes a branching ratio.
The reactions
$\psi'\to \ax{\rm {\rm X}}$, $F_2\to \ax{\rm {\rm X}}$ can go only through chains,
and their branching ratios are
\bea
\Gamma(\psi'\to \ax{\rm {\rm X}}) &=& 
\Gamma(\psi'\to \ax F_2) B(F_2\to {\rm {\rm X}})
+ \Gamma(\psi'\to \tilde F_1\tilde F_2) B(\tilde F_2\to \tilde F_1\ax)
\nonumber\\
&&+2\Gamma(\psi'\to \tilde F_2\tilde F_2)B(\tilde F_2\to \tilde F_1\ax)
B(\tilde F_2\to {\rm X})  \\
\Gamma(F_2\to \ax{\rm {\rm X}}) &=& \Gamma(F_2\to \ax \psi') B(\psi'\to {\rm {\rm X}})
+ \Gamma(F_2\to \tilde F_1\tilde F_2) B(\tilde F_2\to \tilde F_1\ax)
\nonumber\\
&&+2\Gamma(F_2\to \tilde F_2\tilde F_2)B(\tilde F_2\to \tilde F_1\ax)
B(\tilde F_2 \to {\rm X}) \nonumber\\
&&+2\Gamma(F_2\to \psi'\psi')B(\psi'\to \ax{\rm {\rm X}})B(\psi'\to {\rm X})  \nonumber
\,.
\eea
The rates for producing more axions are likely to be  suppressed, 
 because every term in
the analogous expressions containing the product
of more than two or more branching ratios.

Through \eqs{delax}{badef}, these 
 expressions provide the basis for a calculation of
$\delta N_\nu$, as a function of the parameters, if the relative values
of the  initial
flaton densities $n_I$ are known.
Here, we perform the less ambitious task, of trying to identify 
a region of parameter space
allowed by an upper  bound $\delta N_\nu$ of order 1 
 to $0.1$. (Recall that the former bound is roughly the present one
\cite{delnu},
while improvements in the foreseeable future will either tighten the limit
by an order of magnitude, or else detect a nonzero $\delta N_\nu$.)
We shall find that such a region probably does not exist in the 
KSVZ case, but that it certainly does exist in the DFSZ case.

%
%
%

\subsection {Parameter space of KSVZ models}

As  discussed  already,
the flatons $F_{1,2}$ can decay into two gluons,
so let us try to see if the generous requirements
$B_{F_{1,2}}=\Gamma( F_{1,2} \rightarrow {\ax\ax} )/ 
\Gamma( F_{1,2} \rightarrow gg) <3$ [see Eq.~(\ref{delax})]
can be fulfilled in any  region of parameter space.
{}From Eqs.~(\ref{f2aa}), (\ref{f1aa}) and (\ref{glgl}), we get 
\bea
B^{-1}_{F_1}&\simeq&
8\times10^{-4} N_E^2 \(\frac{ \sin \alpha \(x^2+9\)}
{x\(9\cos \alpha-x \sin \alpha\)}  \)^2 \nonumber\\
B^{-1}_{F_2} &\simeq&
8\times10^{-4} N_E^2 \(\frac{ \cos \alpha \(x^2+9\)}
{x\(9\sin \alpha+x \cos \alpha\)}  \)^2 
\eea
for $\alpha_S\(M_{F}\)\simeq 0.1$.  With $\epsilon=+1$, we find that
$B^{-1}_{F_1} < 7\times10^{-4}N_E^2$ for any value of $x$, and 
$B^{-1}_{F_2} < 6\times10^{-4}N_E^2/x^4$ in the limit $x\to 0$.
Therefore, $B_{F_2}$ can be made small enough for $x\sim 0.1$ but
$B_{F_1}$ cannot taking a reasonable value of $N_E$.
%
%
A way out would come from a cosmological evolution of the flatons.  That is,
one could imagine a situation in which 
the flatons oscillate only along the direction of $F_2$
 so that the populations of $F_1$ and $\psi'$ 
(which can decay into $\ax F_1$) are suppressed by the order of 
$10^4$ compared to that of $F_2$.  But this is not probable.
On can find the similar behavior for $\epsilon=-1$ in which case
$B_{F_1}$ can be made small in the limit $x\to 0$.

\subsection {Parameter space of DFSZ models }

In this subsection,
we will try to find a region of parameter space of the DFSZ models.
For this, we consider rather stringent requirement; 
$B_I\lsim 0.1$ for all three flatons and two flatinos. 
According to Section \ref{srelax},
this requirement will ensure that $\delta N_\nu\lsim 0.1$.
Our strategy is to first write down a set of conditions which 
ensure the requirement, and then to identify a region of parameter space
in which these conditions are satisfied.

\subsubsection{The conditions}

We first note that the decay rates calculated in the previous sections are 
functions of the 4 variables $x=v_P/v_Q$, $f/g$, $A_f/\mu$ and $A_g/\mu$ 
disregarding their overall dependence on $F_{\rm PQ}$.
To be as independent as possible of the 
soft supersymmetry breaking parameters 
we will try to make analytic computations on the 
rates of the flaton decays into
Higgs particles, in particular into the lightest 
Higgs boson  ($h^0$) whose mass has an
automatic upper bound of $\sim 140$ GeV \cite{quiros}. 
We will concentrate on the region with
 $\frac{f}{g}$ negative and $ \left|\frac{f}{g} \right|\ll 1$ and $x\gg1$.

To open the decay channels of the flatons into Higgs particles we have,
 in particular, to  impose 
 $M_{\psi'} > M_A>0$ which
requires
 $\;\;\;\frac{A_g}{\mu}< \left|\frac{f}{g} \right|\,x^2 
\;\;\;$  with
\be
\left|\frac{f}{g}\frac{A_f}{\mu} \right|\,x^2>2\,\left
|\frac{A_g}{\mu}+\frac{f}{g}\,x^2 \right| \,.
\ee
The positivity of flaton masses requires
\be
 \left|\frac{ A_f }{ \mu} \right|<  x^2 \left|\frac{ f}{g}\right| \,.
\ee
In case the flatino production rates are sizable, we also  impose
$R_{\tilde{F}_2} \equiv \Gamma(\tilde{F}_2 \to \chi_1 h^0)/
 \Gamma(\tilde{F}_2 \to a \tilde{F}_1)
 > 10$ and 
$M_{\tilde{F}_1}>M_{\chi_1}+M_{h^0}$ to open the decay mode
$\tilde{F}_1 \to \chi_1 h^0$.  These two  conditions give rise to
the restrictions;
\begin{equation}
 \left| f/g\right|<0.02 \,x\, N_{\chi_1}\quad{\rm and}\quad  
  |\frac{f}{g}|> \frac{1}{3\,x}\,. 
\end{equation}
For the latter condition, we required $M_{\tilde F_1} > \mu$.

Then we study, in our limit, the constraints given by the conditions 
$R_{\psi'}\equiv\Gamma(\psi' \to h^0 A^0)/\Gamma(\psi'\to aF_1) > 10$ and 
$R_{F_i}\equiv\Gamma( F_i \to h^0 h^0)/\Gamma(F_i\to a a ) > 10$. 
The ratios $R$ are
\bea
R_{\psi'}&\sim&
 \frac{1}{144}\left(\frac{g}{f}\right)^2
\frac{\mu^2}{A_f^2}\left(\frac{A_g}{\mu}-
6\, \frac{f}{g}\right)^2 \dlabel{condi1} \nonumber
\\
R_{F_1} &\sim&  \frac{1}{ 4 }\frac{\mu^4}{M_{F_1}^4}
(\frac{A_g}{\mu}-2 \,\frac{f}{g}\,x^2+2)^2\dlabel{condi2}
\\ 
R_{F_2} &\sim&  10^{-3}\, x^4 \frac{\mu^4}{M_{F_2}^4} \;(\frac{A_g}{\mu}
+18 \,\frac{f}{g}\,
+2)^2\dlabel{condi3}  \nonumber
\eea
where  
$M_{F_1}^2\sim |\frac{f}{g} \frac{A_f}{\mu}| x^2 \mu^2$  and
$M_{F_2}^2\sim 12\,\frac{f^2}{g^2}  x^2 \mu^2$
for $ \frac{A_f}{\mu}< 12 |\frac{f}{g}|$, 
and $M_{F_1}\leftrightarrow M_{F_2} $ for $ \frac{A_f}{\mu}>
 12 |\frac{f}{g}|$. 

\subsubsection{A viable region of parameter space}
Now we identify a region of parameter space, in which the
above conditions are all satisfied. Recall that we are considering only
the region $x^2\gg1$, $f/g$ negative, and $|f/g|\ll 1$. Within this region,
we consider  the  four regions
\bea
&I)& \left| \frac{A_g}{\mu}\right|>\left| \frac{f}{g}\right|\, x^2
\nonumber\\
&II)& 2<\left| \frac{A_g}{\mu}\right|<\left| \frac{f}{g}\right| \,x^2
\nonumber\\
&III)& \left| \frac{f}{g}\right| <\left| \frac{A_g}{\mu}\right|<2
\nonumber\\
&IV)&   \left| \frac{A_g}{\mu}\right|<\left| \frac{f}{g}\right|
\nonumber
\eea
Depending on $ \frac{A_f}{\mu}< 12 |\frac{f}{g}|$ or $ \frac{A_f}{\mu}>
12 |\frac{f}{g}|$  we define regions ${\it a}$ or region $ {\it b}$.

We find that all of the ${\it a} $ 
regions are forbidden, and so is the $I\!V_b$ region.
The constraints for the other  regions are as follows.
\bea
&I_b& x>14,\; \; \;  A_g<0,\; \; \; 
12 \left|\frac f g\right| < \left|\frac{A_f} \mu \right|
  \nonumber\\
&& \;\; 1< \frac12
\left|\frac {A_f}\mu\right|< \left|\frac f g\right|\frac{x^2} 2
<\left|\frac{A_g}\mu\right| < 2 \(  \left|\frac f g \right| 
\frac {x^2} 2 \)^2 \,.\\
&I\!I_b& x>9,\; \; \; A_g>0,\; \; \;
\left| \frac{f}{g}\right|<3\times 10^{-2} \nonumber\\
&&\;\;
2<\left| \frac{A_f}{\mu}\right|<
 \left| \frac{f}{g}\right| \,x^2,
\;\;\;
2<\left| \frac{A_g}{\mu}\right|<
\left| \frac{f}{g}\right| \,x^2 \,.\\
& I\!I\!I_b& A_g>0,\; \; \;
1 < 2\left|\frac f g\right| x^2,\nonumber\\
&&\;\;\;
2\left|\frac f g\right| < \left|\frac{A_f}\mu\right| \left|\frac f g\right|
 < 2\times 10^{-2},
\;\;\;\;
\left|\frac f g\right| < \left|\frac {A_g}\mu\right| < 2 \,.
\eea
In addition to these conditions, we need to add the original requirements,
that $f/g$ be negative with $|f/g|\ll 1$. (The latter condition
is automatically satisfied in regions $II_b$ and $III_b$, but not
in region $I_b$.) The other original requirement
$x\gg 1$ is automatically satisfied in all three regions.
Note that all  cases one requires
$x^2| f/ g|\gsim 1$.

Within each of these allowed regions, the parameters can take on 
their natural values
$|A_f|\sim |A_g|\sim|\mu|\sim 100\GeV$, $v_P\sim v_Q$
and $|f|\sim |g|\sim 1$, within a factor 10 or so.

\section{Conclusions}
\dlabel{s6}

We have explored the cosmology of a supersymmetric
extension of 
the Standard Model, which has a Peccei-Quinn symmetry broken only by
two `flaton' fields $\phi_P$ and $\phi_Q$. They have two radial 
modes of oscillation and one angular mode (plus the axion), corresponding 
to three flatons with mass of order $100\GeV$, and there are two
flatinos with roughly the same mass. The flatons are the generalizations
of the saxion which appears in non-flaton models, and the flatinos are 
the generalizations of the axino.

The flaton models have the virtue that $\vpq$ is predicted in terms of 
the electroweak scale and the Planck or GUT scale.
In the canonical model we have estimated
$\vpq\sim 10^{10.4\pm 0.9}\GeV$, and in the more complicated ones
$\vpq\gsim 10^{11.8}\GeV$. 

We have assumed that 
$\phi_P$ has a positive effective  mass-squared in the early Universe,
so that thermal inflation occurs, allowing
rather definite 
predictions. The axion is a good
 dark matter candidate in all cases.
(By {\em good}
dark matter candidate we mean one  whose density
is predicted to be roughly in the right ballpark.)
Bearing in mind that PQ strings are produced after thermal inflation,
this simply corresponds to  the received wisdom in the canonical case
where $\vpq\sim 10^{10}\GeV$.
But the axion is also a good candidate when $\vpq$ is bigger,
because
entropy production from flaton decay more than compensates for
the increased axion density before flaton decay.
Besides the axion, the
LSP is also a good dark matter candidate in
the canonical model. (In the other models, the reheat temperature
is too low to thermalize the LSP, which must therefore be unstable.)
In KSVZ models, a 
 third good candidate is the  supermassive particle whose mass
comes mostly from a coupling to one of the flatons.

Our main concern has been with the highly relativistic
axion population that is produced by 
flaton decay. It remains relativistic to the present and therefore
makes no contribution to the dark matter, but it is dangerous
for nucleosynthesis because it is equivalent to very roughly 
$\delta N_\nu\sim 1$ extra neutrino species.
At present, the bound at something like the 2-$\sigma$ level
is $\delta N_\nu<1.8$ for the `high' deuterium nucleosynthesis scenario,
and $\delta N_\nu <0.3$ for the perhaps favored `low' deuterium
scenario. In the foreseeable future one will either have a bound
$\delta N_\nu\lsim 0.1$, or a detection of $\delta N_\nu$.

We have calculated the rates for all relevant channels
and  examined the constraint 
that the energy density of these axions does not upset the
predictions of the standard nucleosynthesis.
We  confirm the earlier conjecture, that
 the KSVZ  case is  probably  ruled out even by the present 
 bound $\delta N_\nu\lsim 1$.
For the DFSZ  case  there are more decay channels.
To evade complicated phase space suppressions 
we concentrate on the decay of the flatons into Higgses,
 as  
the mass of the lightest Higgs boson has naturally
a relatively low upper bound and the mass of the other Higgs boson and flaton 
fields are fixed by the parameters of the flatonic potential itself.
In this way, we have found a region of parameter space, in which $\delta
N_\nu$ will certainly be $\lsim 0.1$, and in which the parameters can
take on their natural values within a factor ten or so.
In its  DFSZ version, the flaton model of PQ symmetry breaking will
be a candidate for explaining a future detection of nonzero $\delta N_\nu$.

An interesting question, lying beyond the present investigation,
is whether the allowed region of parameter space
can be achieved in a supergravity model with 
universal soft parameters.

\newcommand\pl[3]{Phys. Lett.  {\bf #1}, #3 (19#2)}
\newcommand\npb[3]{Nucl. Phys. {\bf B#1}, #3 (19#2)}
\newcommand\pr[3]{Phys. Rep. {\bf #1}, #3 (19#2)}
\newcommand\prl[3]{Phys. Rev. Lett. {\bf #1}, #3 (19#2)}
\newcommand\prd[3]{Phys. Rev. D{\bf #1}, #3 (19#2)}
\newcommand\ptp[3]{Prog. Theor. Phys. {\bf #1}, #3 (19#2)}
\newcommand\rmp[3]{Rev. Mod. Phys. {\bf #1}, #3 (19#2)}
\newcommand\npps[3]{Nucl. Phys. Proc. Suppl. {\bf #1}, #3 (19#2)}

\end{document}